\documentclass[aps,prl,twocolumn,superscriptaddress]{revtex4}
\usepackage{graphicx}
\usepackage{amsmath}
\usepackage{color}

\begin{document}

\title{Spectral collapse in ensembles of meta-molecules}

\author{V. A. Fedotov}
\email{vaf@orc.soton.ac.uk}
\affiliation{Optoelectronics Research Centre, University of
Southampton, SO17 1BJ, UK}

\author{N. Papasimakis}
\affiliation{Optoelectronics Research Centre, University of
Southampton, SO17 1BJ, UK}

\author{E. Plum}
\affiliation{Optoelectronics Research Centre, University of
Southampton, SO17 1BJ, UK}

\author{A. Bitzer}
\affiliation{Department of Molecular and Optical Physics, University of Freiburg, D-79104, Germany}

\author{M. Walther}
\affiliation{Department of Molecular and Optical Physics, University of Freiburg, D-79104, Germany}

\author{P. Kuo}
\affiliation{Institute of Physics, Academia Sinica, Taipei, 11529, Taiwan}

\author{D. P. Tsai}
\affiliation{Department of Physics, National Taiwan University, Taipei 10617, Taiwan}

\author{N. I. Zheludev}
\homepage{www.nanophotonics.org.uk/niz}
\affiliation{Optoelectronics Research Centre, University of
Southampton, SO17 1BJ, UK}

\date{\today}

\begin{abstract} We report on a new collective phenomenon in metamaterials: spectral line collapse with increasing number of the unit cell resonators (meta-molecules). Resembling the behaviour of exotic states of matter, such as Bose-Einstein condensates of excitons and magnons, this new effect is linked to the suppression of radiation losses in periodic arrays. We demonstrate experimentally spectral line collapse at microwave, terahertz and optical frequencies. It emerges as a universal and truly scalable effect underpinned by classical electromagnetic interactions between the excited meta-molecules.
\end{abstract}

\maketitle

The burgeoning field of metamaterials provides unique opportunities to engineer the electromagnetic properties of artificial media and achieve exotic functionalities, such as negative refraction \cite{negref} and cloaking \cite{Cloaking}. Similarly to natural crystals, which are created by arranging individual atoms and molecules in a regular grid, periodic ensembles of subwavelength electromagnetic resonators present an effective medium to an incident with properties not available in natural materials. Here, we study the dependence of the metamaterial properties on the number of meta-molecules in the microwave, THz and optical domain, and demonstrate a new collective phenomenon in metamaterials: in contrast to solid state crystals, where bulk arrangements result in broadening of the individual element spectral line, leading eventually to the formation of absorption bands, regular ensembles of meta-molecules can exhibit the opposite effect, i.e. spectral line collapse. 
\begin{figure}[h!]
\includegraphics[width=0.45\textwidth]{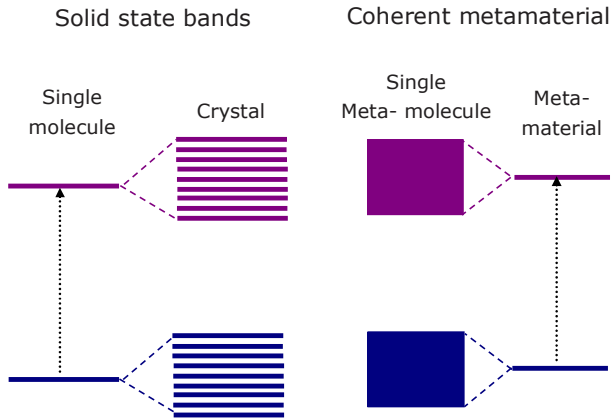}
\caption{(Colour online) Resonant absorption lines of atoms form a
broader absorption band in solids. In coherent metamaterials we see
the opposite trend: resonant lines of large arrays are much narrower
than lines of individual meta-molecules.}
\end{figure}

The reported  phenomenon is characteristic to a novel class of
artificial media, which we call "coherent" metamaterials \cite{CoherentMM} and are characterized by very
strong interactions between the electromagnetically excited
meta-molecules that provide for a low rate of energy loss due to scattering and lead to a high-quality resonant response. An example of a coherent metamaterial is an array of ASRs, where the meta-molecular
excitation corresponds to an oscillating magnetic dipole perpendicular
to the plane of the array that does not interact directly with the
magnetic field of the incident wave, thus creating a nearly
thermodynamically isolated ensemble of strongly interacting
coherent "molecules" with interesting physical properties. To
illustrate this behavior we present a comparison with an
``incoherent" metamaterial: a two-dimensional array formed by
pairs of concentric conducting rings that also supports a
high-quality resonant response. In this case, however, the response of the array is a sum
of the individual meta-molecule responses, rather than a collective
property. 
\begin{figure}[h!]
\includegraphics[width=0.45\textwidth]{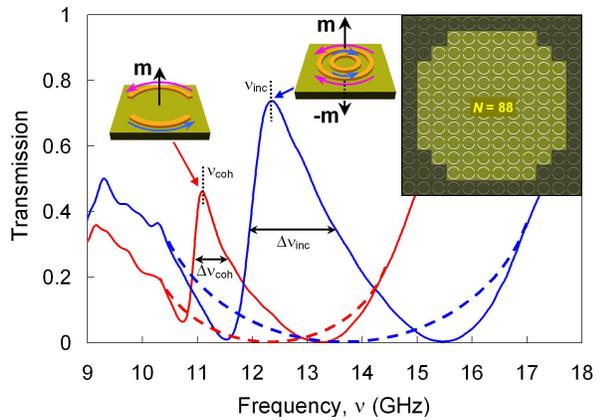}
\caption{(Colour online) Transmission spectra feature trapped-mode resonances of the coherent
(asymmetric split rings, solid red curve, $\nu_{coh}$) and
incoherent (concentric rings, solid blue curve, $\nu_{inc}$)
metamaterial arrays of 88 meta-molecules. Dashed curves indicate
transmission stop bands of the metamaterials in the absence of
trapped-mode excitations. The top right inset shows a sample of 88
meta-molecules (unit cells) of the coherent array exposed through a
metal mask (represented by the darker area). Other insets show the
trapped current modes excited in the unit cells of the coherent
(left) and incoherent (right) metamaterials and the corresponding
induced magnetic dipole moments $m$.}
\end{figure}

The coherent microwave metamaterial was manufactured as a regular planar
array of asymmetric split rings (ASR) etched from a $35~\mu\text{m}$
thick copper layer on a 1.6~mm thick FR4 substrate. The diameter of the ASR was 6~mm  with a line width of 0.4~mm and was split in two segments corresponding to
$140^{\circ}$ and $160^{\circ}$ arcs. The unit cell of $7.5 \times 7.5~\text{mm}^2$ rendered the
arrays non-diffracting at normal incidence for frequencies of up to
40~GHz. In the incoherent metamaterial ASRs were
replaced with pairs of concentric rings. The inner and outer rings
had the diameters of correspondingly $4.50~mm$ and $5.45~mm$, and were both
$0.2~mm$ wide. The transmission measurements were performed in a
microwave anechoic chamber at normal incidence using broadband
linearly polarized horn antennas equipped with collimating lenses
and a vector network analyzer.

Transmission spectra of large coherent and incoherent arrays show
similar resonant features in the form of a broad stop-band split by
a narrower Fano-like transmission peak (see Fig.~2). For the array of ASRs the transmission resonance occurs at around 11~GHz and is
associated with the excitation of anti-symmetric currents
oscillating in the opposite arcs of each split ring (see inset to
Fig.~2). Such a current mode, known as a trapped or closed mode,
yields magnetic dipole moments oscillating synchronously
(coherently) in all meta-molecules along the direction normal to the
plane of the array. The induced magnetic dipoles interact strongly
with one another, while their interaction with the perpendicularly
oriented magnetic field of the incident electromagnetic wave is
forbidden \cite{TrappedMode1}. For the double-ring metamaterial a
similar narrow resonance at 13~GHz corresponds to oscillations of
oppositely directed currents in the inner and outer rings, as shown
in the inset to Fig.~2. Although such current configurations give
rise to magnetic moments, the latter cancel one another and the
total magnetic moment of the meta-molecule and thus interactions
between meta-molecules are negligible \cite{DoubleRings}.

Figure 3 shows the dependencies of the transmission resonance
quality factors, $Q = \nu/ \Delta \nu$, on the total number of
meta-molecules, $N$, that form the arrays. In the experiment, the
number of meta-molecules exposed to the incident wave was controlled
by placing metal masks of different sizes on a large metamaterial
array, as illustrated in the inset to Fig.~2. These masks screened
the peripheral parts of the array, leaving the central part of the
array exposed to electromagnetic radiation. The shape of the masks
was close to circular with a step-like profile of the opening, which
ensured that every unit cell of the array was either fully screened
or exposed. These measurements were performed with 22 different
masks exposing arrays with a total number of unit cells in the range
from 32 to about 700. Experiments with larger arrays were not
practical and unnecessary as already for $N \simeq 700$ the quality
factor saturated. Experiments with smaller arrays were not feasible
as for $N < 32$ diffraction takes its toll on the accuracy of the
data. In the measured range of $N$ the experimental data clearly
indicate that the Q-factor of the coherent metamaterial strongly
depends on the size of the illuminated area, i.e., the total number
of meta-molecules engaged in the interaction with the incident wave.
Indeed, the Q-factor, which measures about 10.5 for the smallest
array ($N=32$), can be seen to increase by almost $70\%$ reaching
17.5 for the full-sized array with $N = 688$. In contrast, the
Q-factor of the incoherent metamaterial appears to be practically
independent of the number of exposed unit cells and remains at
around $Q=9$.
\begin{figure}[h!]
\includegraphics[width=0.45\textwidth]{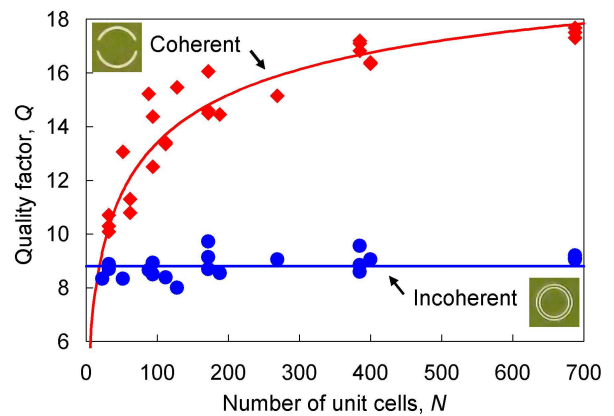}
\caption{(Colour online) Graphs show Q-factors as a function of the
total number of meta-molecules. Experimentally measured data are
represented by points for both coherent (diamonds) and incoherent
(circles) metamaterials. Solid lines present theoretical fits to the
data.}
\end{figure}

To illustrate further the dependence of the resonant response on the
size of the coherent metamaterial, we studied the spatial
distribution of the magnetic field in a single split-ring
meta-molecule and a meta-molecule that is a part of a regular array
containing $N=400$ unit cells. For this we employed a THz near-field
imaging technique with sub-wavelength resolution described in
\cite{BitzerWalther}. The technique enabled accurate mapping of
orthogonal components of the electrical field $\vec{E}$ in the plane of
the unit cell (as illustrated in the inset to Fig.~4a), which was
used to calculate the component of the magnetic field $\vec{H}$ normal to
the plane of the structure via the Maxwell equation
$\vec{\nabla}~\times~\vec{E}=-\partial \vec{B}/\partial t$.
\begin{figure}[h!]
\includegraphics[width=0.45\textwidth]{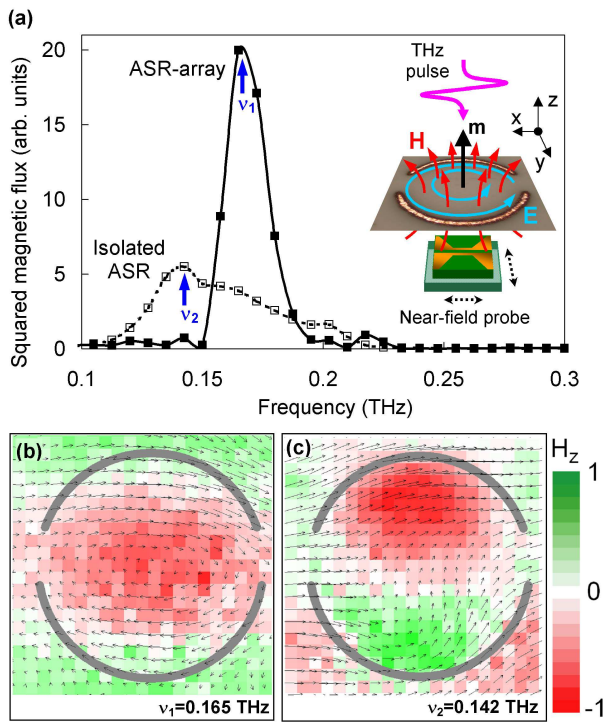}
\caption{(Colour online) Panel (a) shows the spectral dependence
of the squared total magnetic flux amplitude through an ASR for the cases of one isolated ring (empty squares, weak
magnetic response) and a ring located in the middle of a $20 \times
20$ array (filled squares, strong magnetic response). Panels (b) and
(c) present snap-shot maps of the instantaneous component of the
magnetic field normal to the plane of the ring (colour coded) and
the in-plane component of the electric field (arrows) at resonance
frequencies $\nu_1$ and $\nu_2$ correspondingly (see also supplementary videos for temporal evolution of the fields). In order to show both fields at their maxima on the same map the magnetic and electric fields are presented with phase shift of $\pi/4$.}
\end{figure}

In our experiment we observed that at resonance conditions the
spatial distribution of the magnetic field was radically different
for a split-ring placed in the array and an isolated split-ring. For
a ring in the array the magnetic field penetrates the unit cell in
the same direction everywhere within the area enclosed by the ring:
the total magnetic moment of the meta-molecule is at maximum
indicating the excitation of strong anti-symmetric (i.e. ring)
currents in opposite sections of the ring (see Fig.~4b), while the
spectral dependence of the squared amplitude of the magnetic flux
reveals a narrow sharp peak centered at $\nu_1 = 0.165~\text{THz}$
(see Fig.~4a). For an isolated single meta-molecule the net magnetic
flux is also non-zero but it is much lower than for a ring in the
array: the magnetic field is oppositely directed in adjacent
sections of the ring (see Fig.~4c), the total magnetic flux is small
and the anti-symmetric current component is weak. As a result, the
resonant feature at $\nu_2 = 0.142~\text{THz}$ is much less
pronounced and its Q-factor is significantly lower than for rings in
the array indicating considerable damping (see Fig.~4a).

We attribute the observed size-dependent resonant response of the
coherent metamaterial to the existence of strong interactions
between the meta-molecules mediated by magneto-inductive surface
waves \cite{Shamonina}. It follows from symmetry arguments that
magnetic dipoles coherently oscillating perpendicular to the plane
of an infinite (or very large) regular array can re-radiate only in
its plane, which ensures very low scattering losses and therefore a
high Q-factor of the system. Electromagnetic energy in the coherent
array is trapped in the form of standing magneto-inductive surface
waves, which may only scatter on the edges of the array. The
scattering losses grow and the Q-factor diminishes as the array
becomes smaller, and in the limiting case of an isolated
meta-molecule the resonance is extremely weak due to intense energy
losses through magnetic dipole radiation (see also supplementary materials).

The dependence of the Q-factor on the size of the array can be
qualitatively understood from the definition of the Q-factor, $Q = 2
\pi \cdot E / \Delta E$, where $E$ is the energy stored in the array
and $\Delta E$ is the energy dissipated by the array per cycle. Here
$E$ is simply proportional to $N$, while $\Delta E = \Delta E_m +
\Delta E_s$  represents contributions from two loss mechanisms.
$\Delta E_m=\sigma_m \cdot N$ includes material losses, i.e., Ohmic
losses in the metal and dissipation losses in the dielectric
substrate, as well as radiation losses due to an electric dipole
induced in each unit cell, which is controlled by the degree of
asymmetry of the split ring. Losses associated with scattering of
the magneto-inductive waves on the edges of the array, $\Delta E_s$,
are proportional to the length of its perimeter and thus $\Delta
E_s=\sigma_s \cdot \sqrt{N}$. Thus, the phenomenological dependence
of the quality factor on the number of unit cells can be presented
in the form $Q \propto N/( \sigma_m \cdot N + \sigma_s \sqrt{N})$,
which fits the experimental data very well (solid red line in
Fig.~3). In comparison, in the case of the incoherent array of
concentric pairs of rings, the total induced magnetic moment
perpendicular to the plane of the array is small, hence magnetic
dipole interactions and scattering of magneto-inductive waves are
negligible ($\sigma_m \cdot N \gg \sigma_s \sqrt{N}$). The second
term in the denominator can be disregarded making the Q-factor
independent of the size of the array, which is in complete agreement
with our experimental observations (solid blue line in Fig.~3).

The dependence of the resonant response on the array's size results
from truly classical interactions between the excited states of the
meta-molecules and therefore shall be universal and truly scalable
with wavelength. Since the characteristic size of the array at which
substantial broadening of the resonance is seen is given by $N =
(\sigma_s / \sigma_m)^2$, with increase of Joule losses the size
dependent effect will be seen in smaller arrays, i.e., in an array
with higher Joule losses the coherent state will be formed by a
smaller number of meta-molecules.

This is exactly what we observed in our experiments in the optical
part of the spectrum, where Joule losses are dominant. Here we
studied transmission of square asymmetric split ring slit arrays of different
sizes scaled down 15,000 times relative to the microwave sample.  The ring
slits (complimentary to split-rings) were cut from a 55 nm thick gold film supported by a $500~\mu\text{m}$ silica substrate using focused ion beam milling. The transmission measurements were performed using a
microscope-based spectrophotometer by CRAIC Technologies. Because of
the \emph{negative} configuration the size of the illuminated area in
transmission was naturally controlled by the apertures of the
arrays, while unstructured square windows of the same size were used
as reference. We studied arrays of five different sizes
containing from 16 to 144 meta-molecules (see Fig.~5a - 5b). Such
photonic metamaterials exhibited a trapped-mode resonance at around
1000~nm, which, in full accordance with Babinet's principle, was
seen not as a transmission peak (as for a \emph{positive} array, see
Fig.~2), but as a dip in transmission for the largest array of $12
\times 12$ unit cells (see inset to Fig.~5c). For smaller arrays the
resonance gradually became broader and shallower and, as shown in
Fig.~5c, completely disappeared for $N=16$, which suggests that the
most dramatic change occurs when the number of unit cells drops
below 49.
\begin{figure}[h!]
\includegraphics[width=0.45\textwidth]{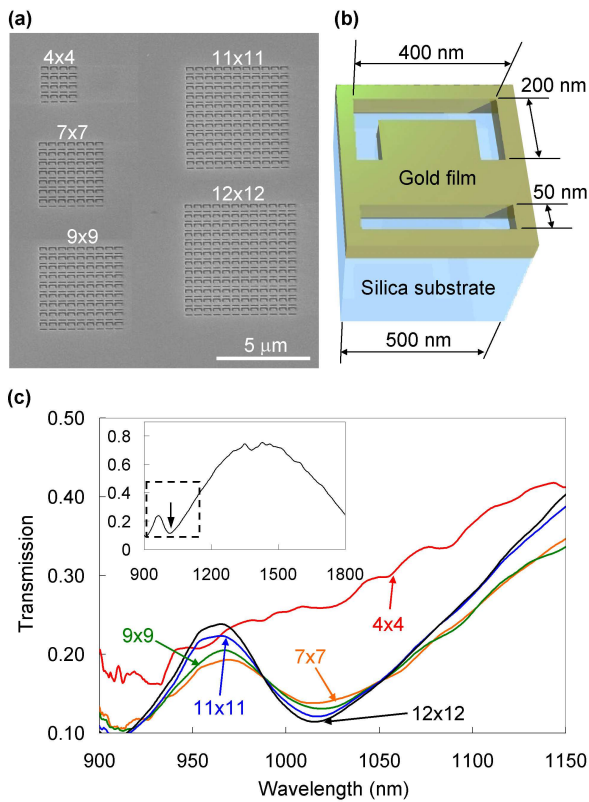}
\caption{(Colour online)  Panel (a) shows SEM images of coherent
metamaterial arrays of  different sizes. Panel (b) presents a
schematic of the metamaterial unit cell. Panel (c) shows
transmission spectra of the metamaterial arrays in the vicinity of
the trapped-mode resonance. The inset presents the transmission
spectrum of the $12 \times 12$ array over a much wider wavelength
range; a dashed box indicates the spectral domain that is covered by
the main plot, while an arrow marks the position of the trapped-mode
resonance.}
\end{figure}

The results presented above illustrate that larger arrays of
coherent metamaterials exhibit narrower resonances. They also show
that in samples with smaller losses more meta-molecules 
are engaged in forming the coherent state: the Q-factor starts
reducing dramatically below about 200 meta-molecules in the
microwave array, while in the lossy optical case we saw a
significant reduction of Q-factor below 49 meta-molecules. We argue
that the number of molecules forming the coherent response will
increase as Joule and dielectric losses are compensated. This is
relevant to the recent observation of loss compensation in a
coherent photonic metamaterial \cite{QD_ASR}, which paves the way to
an intriguing opportunity of creating a lasing spaser, a coherent
source of optical radiation fuelled by coherent plasmonic
oscillations in individual metamaterial resonators of a coherent
array. In the latter case the gain substrate supporting the rings
would be the source of energy \cite{LSPASER}.

In conclusion, we would like to
point out that the spectral collapse of the metamaterial resonance
with increasing number of meta-molecules has intriguing
phenomenological similarity to the narrowing of
luminescence/scattering spectra upon formation of the
Bose-Einstein condensates (BEC) of exciton and magnon in solids
\cite{BEC1,BEC2,BEC3}. The latter is observed when concentration
of the bosonic quasi-particles in equilibrium exceeds a certain
critical value and most of the particles are statistically forced
to occupy the ground level forming a stable coherent eigenstate.
In the case of the excitons, for example, the narrow emission line
of the condensate corresponds to the binding energy of the
quasi-particles and results from their recombination in the
condensate. The analogy between these two phenomena becomes
apparent since the ground state of the metamaterial system is
achieved through de-excitation of the low scattering (i.e.
trapped) electromagnetic mode of interacting meta-molecules.
Moreover, as in the BEC, the collapse of the resonant line is most
rapid when the number of meta-molecules reaches a certain critical
value ($N_c \approx 100$ in the case of the microwave
metamaterial).


\begin{thebibliography}{12}
\expandafter\ifx\csname natexlab\endcsname\relax\def\natexlab#1{#1}\fi
\expandafter\ifx\csname bibnamefont\endcsname\relax
  \def\bibnamefont#1{#1}\fi
\expandafter\ifx\csname bibfnamefont\endcsname\relax
  \def\bibfnamefont#1{#1}\fi
\expandafter\ifx\csname citenamefont\endcsname\relax
  \def\citenamefont#1{#1}\fi
\expandafter\ifx\csname url\endcsname\relax
  \def\url#1{\texttt{#1}}\fi
\expandafter\ifx\csname urlprefix\endcsname\relax\def\urlprefix{URL }\fi
\providecommand{\bibinfo}[2]{#2}
\providecommand{\eprint}[2][]{\url{#2}}

\bibitem[{\citenamefont{Smith and andM. C.~K.~Wiltshire}(2004)}]{negref}
\bibinfo{author}{\bibfnamefont{D.~R.} \bibnamefont{Smith}} \bibnamefont{and}
  \bibinfo{author}{\bibfnamefont{J.~B.~P.} \bibnamefont{andM.
  C.~K.~Wiltshire}}, \bibinfo{journal}{Science} \textbf{\bibinfo{volume}{305}},
  \bibinfo{pages}{788} (\bibinfo{year}{2004}).

\bibitem[{\citenamefont{Schurig et~al.}(2006)\citenamefont{Schurig, Mock,
  Justice, Cummer, Pendry, Starr, and Smith}}]{Cloaking}
\bibinfo{author}{\bibfnamefont{D.}~\bibnamefont{Schurig}},
  \bibinfo{author}{\bibfnamefont{J.~J.} \bibnamefont{Mock}},
  \bibinfo{author}{\bibfnamefont{B.~J.} \bibnamefont{Justice}},
  \bibinfo{author}{\bibfnamefont{S.~A.} \bibnamefont{Cummer}},
  \bibinfo{author}{\bibfnamefont{J.~B.} \bibnamefont{Pendry}},
  \bibinfo{author}{\bibfnamefont{A.~F.} \bibnamefont{Starr}}, \bibnamefont{and}
  \bibinfo{author}{\bibfnamefont{D.~R.} \bibnamefont{Smith}},
  \bibinfo{journal}{Science} \textbf{\bibinfo{volume}{314}},
  \bibinfo{pages}{977} (\bibinfo{year}{2006}).

\bibitem[{\citenamefont{Papasimakis et~al.}(2009)\citenamefont{Papasimakis,
  Fedotov, Fu, Tsai, and Zheludev}}]{CoherentMM}
\bibinfo{author}{\bibfnamefont{N.}~\bibnamefont{Papasimakis}},
  \bibinfo{author}{\bibfnamefont{V.~A.} \bibnamefont{Fedotov}},
  \bibinfo{author}{\bibfnamefont{Y.~H.} \bibnamefont{Fu}},
  \bibinfo{author}{\bibfnamefont{D.~P.} \bibnamefont{Tsai}}, \bibnamefont{and}
  \bibinfo{author}{\bibfnamefont{N.~I.} \bibnamefont{Zheludev}},
  \bibinfo{journal}{Phys. Rev. B} \textbf{\bibinfo{volume}{80}},
  \bibinfo{pages}{041102(R)} (\bibinfo{year}{2009}).

\bibitem[{\citenamefont{Fedotov et~al.}(2007)\citenamefont{Fedotov, Rose,
  Prosvirnin, Papasimakis, and Zheludev}}]{TrappedMode1}
\bibinfo{author}{\bibfnamefont{V.~A.} \bibnamefont{Fedotov}},
  \bibinfo{author}{\bibfnamefont{M.}~\bibnamefont{Rose}},
  \bibinfo{author}{\bibfnamefont{S.~L.} \bibnamefont{Prosvirnin}},
  \bibinfo{author}{\bibfnamefont{N.}~\bibnamefont{Papasimakis}},
  \bibnamefont{and} \bibinfo{author}{\bibfnamefont{N.~I.}
  \bibnamefont{Zheludev}}, \bibinfo{journal}{Phys. Rev. Lett.}
  \textbf{\bibinfo{volume}{99}}, \bibinfo{pages}{147401}
  (\bibinfo{year}{2007}).

\bibitem[{\citenamefont{Papasimakis et~al.}(in press)\citenamefont{Papasimakis,
  Fu, Fedotov, Prosvirnin, Tsai, and Zheludev}}]{DoubleRings}
\bibinfo{author}{\bibfnamefont{N.}~\bibnamefont{Papasimakis}},
  \bibinfo{author}{\bibfnamefont{Y.~H.} \bibnamefont{Fu}},
  \bibinfo{author}{\bibfnamefont{V.~A.} \bibnamefont{Fedotov}},
  \bibinfo{author}{\bibfnamefont{S.~L.} \bibnamefont{Prosvirnin}},
  \bibinfo{author}{\bibfnamefont{D.~P.} \bibnamefont{Tsai}}, \bibnamefont{and}
  \bibinfo{author}{\bibfnamefont{N.~I.} \bibnamefont{Zheludev}},
  \bibinfo{journal}{Appl. Phys. Lett.}  \textbf{\bibinfo{volume}{94}}, \bibinfo{pages}{211902} (\bibinfo{year}{2009}).

\bibitem[{\citenamefont{Bitzer et~al.}(2009)\citenamefont{Bitzer, Merbold,
  Thoman, Feurer, Helm, and Walther}}]{BitzerWalther}
\bibinfo{author}{\bibfnamefont{A.}~\bibnamefont{Bitzer}},
  \bibinfo{author}{\bibfnamefont{H.}~\bibnamefont{Merbold}},
  \bibinfo{author}{\bibfnamefont{A.}~\bibnamefont{Thoman}},
  \bibinfo{author}{\bibfnamefont{T.}~\bibnamefont{Feurer}},
  \bibinfo{author}{\bibfnamefont{H.}~\bibnamefont{Helm}}, \bibnamefont{and}
  \bibinfo{author}{\bibfnamefont{M.}~\bibnamefont{Walther}},
  \bibinfo{journal}{Opt. Express} \textbf{\bibinfo{volume}{17}},
  \bibinfo{pages}{3826} (\bibinfo{year}{2009}).

\bibitem[{\citenamefont{Shamonina et~al.}(2002)\citenamefont{Shamonina,
  Kalinin, Ringhofer, and Solymar}}]{Shamonina}
\bibinfo{author}{\bibfnamefont{E.}~\bibnamefont{Shamonina}},
  \bibinfo{author}{\bibfnamefont{V.~A.} \bibnamefont{Kalinin}},
  \bibinfo{author}{\bibfnamefont{K.~H.} \bibnamefont{Ringhofer}},
  \bibnamefont{and} \bibinfo{author}{\bibfnamefont{L.}~\bibnamefont{Solymar}},
  \bibinfo{journal}{J. Appl. Phys.} \textbf{\bibinfo{volume}{99}},
  \bibinfo{pages}{6252} (\bibinfo{year}{2002}).

\bibitem[{\citenamefont{Plum et~al.}(2009)\citenamefont{Plum, Fedotov, Kuo,
  Tsai, and Zheludev}}]{QD_ASR}
\bibinfo{author}{\bibfnamefont{E.}~\bibnamefont{Plum}},
  \bibinfo{author}{\bibfnamefont{V.~A.} \bibnamefont{Fedotov}},
  \bibinfo{author}{\bibfnamefont{P.}~\bibnamefont{Kuo}},
  \bibinfo{author}{\bibfnamefont{D.~P.} \bibnamefont{Tsai}}, \bibnamefont{and}
  \bibinfo{author}{\bibfnamefont{N.~I.} \bibnamefont{Zheludev}},
  \bibinfo{journal}{Opt. Express} \textbf{\bibinfo{volume}{17}},
  \bibinfo{pages}{8548} (\bibinfo{year}{2009}).

\bibitem[{\citenamefont{Zheludev et~al.}(2008)\citenamefont{Zheludev,
  Prosvirnin, Papasimakis, and Fedotov}}]{LSPASER}
\bibinfo{author}{\bibfnamefont{N.~I.} \bibnamefont{Zheludev}},
  \bibinfo{author}{\bibfnamefont{S.~L.} \bibnamefont{Prosvirnin}},
  \bibinfo{author}{\bibfnamefont{N.}~\bibnamefont{Papasimakis}},
  \bibnamefont{and} \bibinfo{author}{\bibfnamefont{V.~A.}
  \bibnamefont{Fedotov}}, \bibinfo{journal}{Nat. Photonics}
  \textbf{\bibinfo{volume}{2}}, \bibinfo{pages}{351} (\bibinfo{year}{2008}).

\bibitem[{\citenamefont{Lin and Wolfe}(1993)}]{BEC1}
\bibinfo{author}{\bibfnamefont{J.~L.} \bibnamefont{Lin}} \bibnamefont{and}
  \bibinfo{author}{\bibfnamefont{J.~P.} \bibnamefont{Wolfe}},
  \bibinfo{journal}{Phys. Rev. Lett.} \textbf{\bibinfo{volume}{71}},
  \bibinfo{pages}{1222} (\bibinfo{year}{1993}).

\bibitem[{\citenamefont{Demokritov et~al.}(2006)\citenamefont{Demokritov,
  Demidov, Dzyapko, Melkov, Serga, Hillebrands, and Slavin}}]{BEC2}
\bibinfo{author}{\bibfnamefont{S.~O.} \bibnamefont{Demokritov}},
  \bibinfo{author}{\bibfnamefont{V.~E.} \bibnamefont{Demidov}},
  \bibinfo{author}{\bibfnamefont{O.}~\bibnamefont{Dzyapko}},
  \bibinfo{author}{\bibfnamefont{G.~A.} \bibnamefont{Melkov}},
  \bibinfo{author}{\bibfnamefont{A.~A.} \bibnamefont{Serga}},
  \bibinfo{author}{\bibfnamefont{B.}~\bibnamefont{Hillebrands}},
  \bibnamefont{and} \bibinfo{author}{\bibfnamefont{A.~N.}
  \bibnamefont{Slavin}}, \bibinfo{journal}{Nature}
  \textbf{\bibinfo{volume}{443}}, \bibinfo{pages}{430} (\bibinfo{year}{2006}).

\bibitem[{\citenamefont{Chumak et~al.}(2009)\citenamefont{Chumak, Melkov,
  Demidov, Dzyapko, Safonov, and Demokritov}}]{BEC3}
\bibinfo{author}{\bibfnamefont{A.~V.} \bibnamefont{Chumak}},
  \bibinfo{author}{\bibfnamefont{G.~A.} \bibnamefont{Melkov}},
  \bibinfo{author}{\bibfnamefont{V.~E.} \bibnamefont{Demidov}},
  \bibinfo{author}{\bibfnamefont{O.}~\bibnamefont{Dzyapko}},
  \bibinfo{author}{\bibfnamefont{V.~L.} \bibnamefont{Safonov}},
  \bibnamefont{and} \bibinfo{author}{\bibfnamefont{S.~O.}
  \bibnamefont{Demokritov}}, \bibinfo{journal}{Phys. Rev. Lett.}
  \textbf{\bibinfo{volume}{102}}, \bibinfo{pages}{187205}
  (\bibinfo{year}{2009}).

\end{thebibliography}
\end{document}